\documentclass[a4paper,aps,pre,twocolumn,groupedaddress,showkeys,showpacs,floats,floatats,floatfix,longbibliography]{revtex4-1}
\usepackage[utf8]{inputenc}
\usepackage[english]{babel}
\usepackage{graphicx}
\usepackage{dcolumn} 
\usepackage{bm}
\usepackage{latexsym}
\usepackage{natbib}
\usepackage{mathrsfs}
\usepackage{amssymb}
\usepackage{amsmath, mathtools}
\usepackage{amscd}
\usepackage{color}
\usepackage{pifont}
\usepackage{pstricks,pst-node,pst-text,pst-3d}
\usepackage{verbatim}
\usepackage{soul}
\usepackage[T1]{fontenc}
\usepackage{hyperref}
\hypersetup{
  colorlinks   = true, %colors links instead of ugly boxes
  urlcolor     = black, %color for external hyperlinks
  linkcolor    = red, %color of internal links
  citecolor    = blue %color of citations
}
\definecolor{magenta}{rgb}{1.0, 0.0, 1.0}
\definecolor{orange}{rgb}{0.98, 0.6, 0.01}
\definecolor{brown}{rgb}{0.59, 0.29, 0.0}
\definecolor{aquamarine}{rgb}{0.5, 1.0, 0.83}
\definecolor{blue-violet}{rgb}{0.54, 0.17, 0.89}
\definecolor{verde}{rgb}{0.04, 0.6, 0.02}
\definecolor{cinza}{rgb}{0.6, 0.6, 0.6}

\newcommand{\cbeims}[1]{{\color{black}#1}}
\newcommand{\xxr}[1]{{\color{red}#1}}
\newcommand{\rms}[1]{{\color{black}#1}}
\newcommand{\xxb}[1]{{\color{blue}#1}}

\newcommand{\xxg}[1]{{\color{verde}#1}}

\newrgbcolor{Red}{1.0 0.0 1.0}
\begin{document}
\title{Strong correlations between power-law growth of COVID-19 in four\\
 continents and the inefficiency of soft quarantine strategies \\
}
\author{Cesar Manchein$^1$}
\email{cesar.manchein@udesc.br}

\author{Eduardo L.~Brugnago$^{2}$}
\email{elb@fisica.ufpr.br}

\author{Rafael M.~da Silva$^{2}$}
\email{rmarques@fisica.ufpr.br}

\author{Carlos F.O.~Mendes$^3$}
\email{cfabio.mendes@gmail.com}
\author{Marcus W.~Beims$^{2}$}
\email{mbeims@fisica.ufpr.br}
\affiliation{$^1$Departamento de F\'\i sica, Universidade do Estado de
  Santa Catarina, 89219-710 Joinville, SC, Brazil}
\affiliation{$^2$Departamento de F\'\i sica, Universidade eFederal do
  Paran\'a, 81531-980 Curitiba, PR, Brazil}
\affiliation{$^3$Escola Normal Superior, Universidade do Estado do Amazonas, 
69050-010 Manaus, AM,  Brazil}
\date{\today}
%
%===============================================================================
\begin{abstract}
In this work we analyse the growth of the cumulative number of
confirmed infected cases by the COVID-19 until March
{27$^{\mbox{\footnotesize th}}$,} 2020, from countries of Asia,
Europe, North and South America. Our results show (i) that power-law
growth is observed  for all countries; (ii) by using the
  distance correlation, that the power-law  curves between countries
are statistically highly correlated,  suggesting the
universality  of such curves around the World; and (iii) that
{\it soft} quarantine strategies are  inefficient to flatten
the growth curves. Furthermore, we present a model and strategies
which allow the government to reach the flattening of the
power-law curves. We found that, besides the social distance
of individuals, of well known relevance, the strategy of
{\it  identifying} and {\it isolating} infected individuals in a
large daily rate, can help  to flatten the power-laws. These are
essentially the strategies used in the Republic of Korea.  The high
correlation   between the power-law curves of different countries
strongly indicate that the government  containment measures
can be applied with success around the whole World. These measures
must be  scathing and applied as soon as possible. 
\end{abstract}
%===============================================================================
%
%\pacs{05.45.Ac,05.45.Pq}
\maketitle
%===============================================================================

{\bf Since the identification of a novel coronavirus (COVID-19) in Wuhan, China, in 
December 2019, the virus kept spreading around the World. One of the most remarkable 
characteristics of COVID-19 is its high infectivity, resulting in a global pandemic. In 
this complex scenario, tasks like protect the people from the infection and the global 
economy may be considered two of the greater challenges nowadays. In order to 
improve our knowledge about the COVID-19 and its behavior in different countries over the 
World, we exhaustively explore the real time-series of cumulative number of the confirmed 
infected cases by the COVID-19 in the last months until March {27$^{\mbox{\footnotesize 
th}}$,} 2020. In our analysis we considered countries of Asia, Europe, North and South 
America. Our main findings clearly show the existence of a well established power-law 
growth and a strong correlation between power-law curves obtained for different countries. 
These two observations strongly suggest an universal behavior of such curves around the World. 
To improve our analysis, we use a model with six autonomous ordinary differential equations, 
based on the well-known SEIR (Susceptible-Exposed-Infectious-Recovered) epidemic model 
(considering quarantine procedures) to propose efficient strategies which allow the 
government to increase the flattening of the power-law curves. Additionally, we 
also show that soft measures of quarantine are inefficient to flatten the growths curves.} 

\section{Introduction}
\label{intro}
The astonishing increase of positively diagnosed cases due to COVID-19 has called the 
attention of the whole World, including researchers of many areas and governments. It 
urges \cite{puevo20} to find explanations for the already known data and models which may  
allow us to better understand the evolution of the viruses. Such explanations and models 
can hopefully be used to implement social policies and procedures to decrease the number 
of infections and deaths. Time urges to avoid economic and social catastrophes. 

In general, the average reproductive number $R_0$, which gives the number of secondary 
infected individuals generated by a primary infected individual, is the key quantity 
which determines the dynamical evolution of the epidemic \cite{vas06}. Usually, for 
values $R_0 < 1$, the number of new infected individuals decreases exponentially. For 
$1 < R_0 < \infty$ this number increases exponentially \cite{vas06, Rfactor}. However, 
nature is full of surprises and there are plenty of cases for which the exponential 
behavior is substituted by power-law \cite{gosma11} and are related to branching 
processes with diverging reproductive number \cite{vas06}, scale free networks and 
small worlds \cite{watts04}. It was already suggested in the literature that the 
COVID-19 growth might be a small world \cite{ray20}. This is in agreement with recent 
results \cite{singer20} suggesting that for many countries around the World the 
COVID-19 growth has the tendency to follow the power-law.

In fact, recent analysis regarding the behavior of the COVID-19 in China demonstrated a 
power-law $t^{\mu}$ growth of infected cases \cite{maier20}. Authors found exponents 
around $\mu=2.1\pm 0.3$, which do not vary very much for different provinces in China. 
This suggests that socio-economical differences, local geography, differences in 
containment strategies, and heterogeneities essentially affect the {\it value} of the 
exponent $\mu$, but not the qualitative behavior. A model of coupled differential equations, 
which includes quarantine and isolations effects, was used by the authors to match real 
data. Power-law growths for China were obtained also in another study and a possible relation 
to fractal kinetics and graph theory is discussed \cite{ziff20}. 

In line to the above last week publications, the present work analyzes the time-series 
evolution of the COVID-19 for the following countries: Brazil, China, France, Germany, 
Italy, Japan, Republic of Korea, Spain, and United States of America (USA). In all cases 
we observe a power-law increase for the positive detected individuals, where the exponent 
$\mu$ changes for different countries. In addition to the power-law behavior we also computed 
the Distance Correlation (DC) \cite{Szek07} between pairs of countries. The DC is able to 
detect nonlinear correlations between data \cite{CFMWB, Carlos19}. We show that power-law 
data are highly correlated between all analyzed countries. This strongly suggest that 
government strategies to flatten the power-law growth, valid for one country, can be 
successfully applied to other countries and continents. Furthermore, a model of Ordinary 
Differential Equations (ODEs) is proposed and some strategies to flatten the power-law  
curves are discussed using the numerical simulations.

The paper is divided as follows. Section~\ref{data} presents the power-law growth of 
confirmed infected cases of COVID-19 and the DC between pairs of countries is determined. 
Section~\ref{model} discusses numerical results using the proposed model showing many
strategies to flatten the power-law growth. In Section~\ref{con} we summarize our results. 

\section{Real data analysis}
\label{data}

\subsection{Power-law growths}

Figure~\ref{rat} displays data of the cumulative number of confirmed positive infected 
cases by COVID-19 of nine countries as a function of the days. The analyzed countries are 
(in alphabetic order): Brazil, China, France, Germany, Italy, Japan, Republic of Korea, 
Spain, and USA. Data were collected from the situation reports published daily by the 
World Health Organization (WHO) \cite{OMS}. We notice that the values in the vertical axis 
in Fig.~\ref{rat} change for different countries. Initial data regarding the incubation 
time were discarded since they do not contribute to the essential results discussed here. 
Black-continuous curves are the corresponding fitting curves $\alpha+\beta\, t^{\,\mu}$, 
where $t$ is the time given in days, $\alpha$, $\beta$, and $\mu$ are parameters. The 
insets in all plots show the data in the log-log scale. Straight lines in the log-log plot 
represent power-law growth. The fact that the growth increases as a power-law is good news
since it increases  slower than the exponential one. However, that is not good enough.

\begin{table}[!b]
  \caption{Details about the parameters of the fitting curves for the power-law behavior 
    $\alpha+\beta\, t^{\mu}$ shown in Fig.~\ref{rat}.}
  \begin{tabular}{|c|c|c|c|}\hline
    Country & $\alpha$ & $\beta$ & $\mu$ \\ \hline
    USA     & 0        & 0.009   & 4.994 $\pm$ 0.216 \\ \hline
    Germany & 0        & 0.223   & 3.734 $\pm$ 0.107 \\ \hline
    Spain   & 308      & 0.386   & 3.686 $\pm$ 0.037 \\ \hline
    France  & 280      & 0.467   & 3.341 $\pm$ 0.031 \\ \hline
    Italy   & 0        & 2.868   & 2.934 $\pm$ 0.040 \\ \hline
    China   & 98       & 24.013  & 2.492 $\pm$ 0.020 \\ \hline
    Brazil  & 59       & 18.450  & 1.971 $\pm$ 0.054 \\ \hline
    Japan   & 112      & 3.107   & 1.685 $\pm$ 0.034 \\ \hline
    Republic of Korea & 0  & 62.574 & 1.670 $\pm$ 0.065 \\ \hline
  \end{tabular}
  \label{T1}
\end{table}

The regimes with power-law growth are the most relevant to be discussed since they provide 
essential information of what is expected for the future and possible attitudes needed to 
flatten the curves. The exponent $\mu$ changes for distinct countries and the complete 
fitting parameters are given in Table~\ref{T1}. Results in Table~\ref{T1} are presented in 
decreasing order of the exponent $\mu$. USA [Fig.~\ref{rat}(a)] has by far the largest 
exponent and therefore became already the country with epidemic records. Even though 
Germany [Fig.~\ref{rat}(b)] reported a small number of deaths, it has the second large 
exponent, followed by Spain [Fig.~\ref{rat}(c)], France [Fig.~\ref{rat}(d)] and Italy 
[Fig.~\ref{rat}(e)], in this order. China [Fig.~\ref{rat}(f)], Brazil [Fig.~\ref{rat}(g)], 
Japan [Fig.~\ref{rat}(h)], and Republic of Korea [Fig. \ref{rat}(i)], in this order, are 
the last in the list. In the case of China and Republic of Korea the power-laws are more 
clear due to the number of available  data. For these two countries a flatten is observed 
after the power-law. The jump observed after 30 days in China data are due to a change in 
the counting procedure of infected cases (see the situation report on February 
17$^{\mbox{\footnotesize  th}}$, 2020, in Ref. \cite{OMS}). Republic of Korea, on the 
other hand, focused on identifying infected patients immediately and isolating them to
interrupt transmission \cite{Korea}. It is interesting to note that for Japan, another 
country that adopted similar measures, we obtained a similar value for $\mu$.
 
\begin{widetext}
  $\quad$
  \begin{figure}[!t]
    \centering
    \includegraphics[width=0.95\columnwidth]{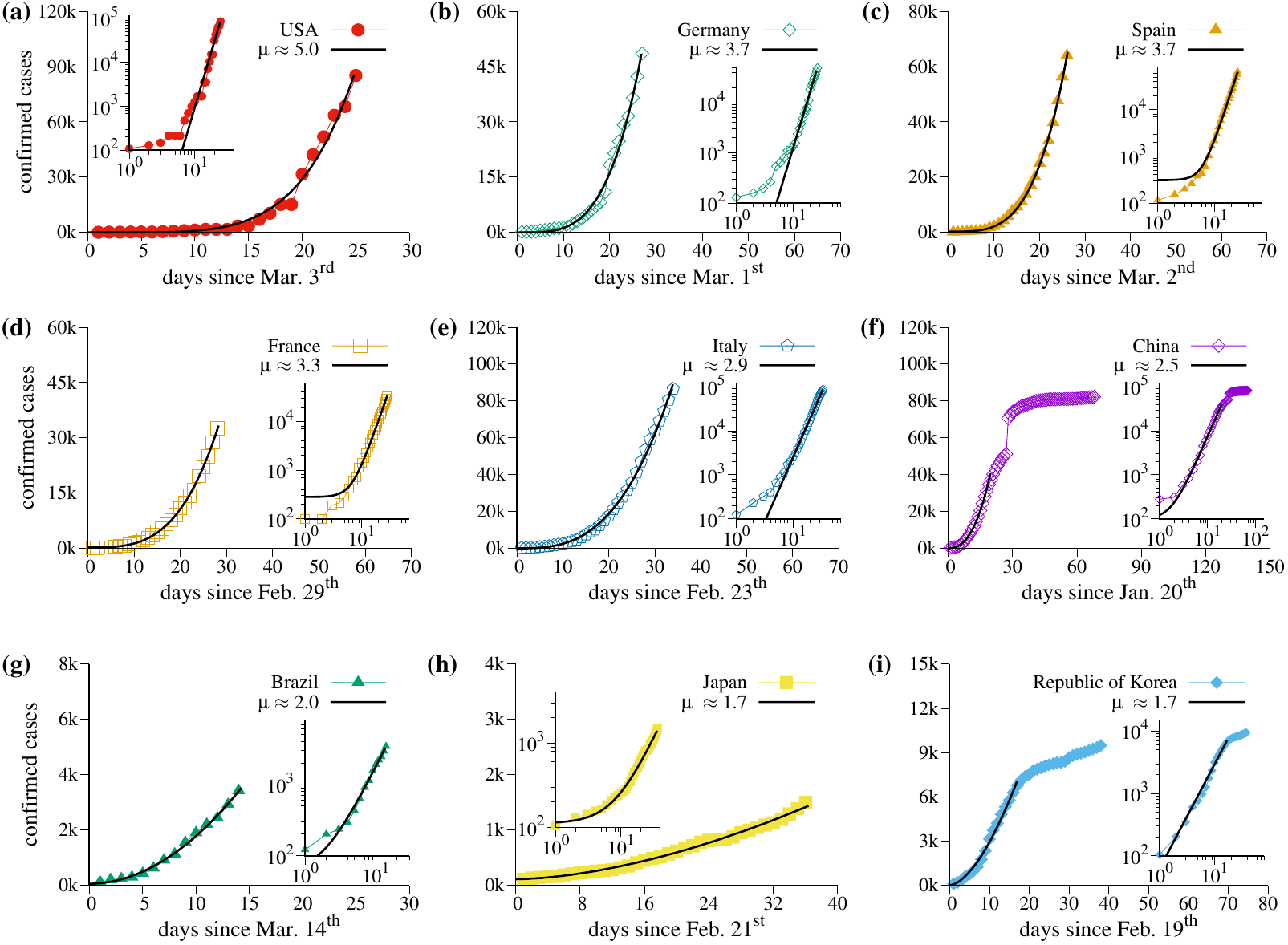}
    \caption{Cumulative number of confirmed infected cases by COVID-19 as a function of 
      time for (a) USA, (b) Germany, (c) Spain, (d) France, (e) Italy,
      (f) China, (g) Brazil, (h) Japan and (i) Republic of Korea,
      excluding days with less than 100  infected. The
      black-continuous curves represent the function $\alpha + \beta
      t^{\mu}$  that fit the time-series, and the parameters $\alpha$,
      $\beta$, and $\mu$ for each  country are described in
      Table~\ref{T1}.} 
    \label{rat}
  \end{figure}
\end{widetext}

The most desired behavior is that the exponent $\mu$ becomes smaller leading to a 
flattening of the curves. But this is apparently not that easy. Besides USA and Germany, 
which have a distinct inclination in the beginning of their power-laws, and China and 
Republic of Korea, which are stabilizing the epidemic spread, for all other countries
the growth remains strictly on the fitted curve and $\mu$ essentially does not change in 
time. In Sec.~\ref{model} we discuss some possibilities to flatten the power-laws.  

\subsection{Distance Correlation between countries}

The power-law observed in all cases from Fig.~\ref{rat} are certainly not a coincidence, 
but a consequence of virus propagation in scale free systems. To quantify the relation 
between the power-law growth we use the DC, which is a statistical measure of dependence 
between random vectors \cite{Szek07, Szek09, Szek12, Szek13, Szek14}. Please do not 
confuse the word distance with the geographical distance between the analyzed countries. 
The most relevant characteristics of DC is that it will be zero if and only if the data 
are independent and equal to one for maximal correlation between data. Details about the 
definition of DC are given in Appendix~\ref{apa}.

Figure \ref{DC} presents specific results for the DC calculated between some selected 
countries, namely Brazil, Italy, Japan, and USA. Italy were chosen due to their relevance 
in Europe, relevance regarding to typical data of the virus. USA was chosen for being 
nowadays the top affected country and Brazil and Japan representing distinct
continents and distinct epidemic containment measures. Thus, we compute the DC 
between four continents. Figures \ref{DC}(a), \ref{DC}(b), \ref{DC}(c),
\ref{DC}(g), \ref{DC}(h), and \ref{DC}(i) are the cumulative number of confirmed 
cases in each country, as in Fig.~\ref{rat}, but considering data since the first day the 
infections were reported. In these curves we clearly see the initial plateaus due to the 
incubation time. After the plateaus, a qualitative change to a power-law growth (the same 
from Fig.~\ref{rat}) occurs. The time for which the qualitative change occurs is distinct 
for each country. 

Figures \ref{DC}(d), \ref{DC}(e), \ref{DC}(f), \ref{DC}(j), \ref{DC}(k), and \ref{DC}(l) 
display the corresponding DC calculated between the countries. Results show that DC
between the curves is relatively high in the beginning. The lowest values are
obtained for the DC between Brazil and Italy, in Fig. \ref{DC}(d), and for the DC between 
Italy and USA, shown in Fig. \ref{DC}(k), both cases around DC $=0.4$. The DC decays 
substantially when the power-law starts in one country but not in the other. The
exception is between Japan and USA. After some days, when both countries reach the 
power-law behavior, the values of DC become very close to $1$. Thus, they are {\it 
highly} correlated besides distinct exponents $\mu$. Furthermore, the DC is not 
necessarily related to the exponent $\mu$. One example can be mentioned. Even though 
USA has the largest exponent and Japan the lowest one (considering the error in Table 
\ref{T1}), they are highly correlated. Besides that, even though there are not many data
available for Brazil, it seems to become more and more correlated with Italy and
Japan.

\begin{widetext}
$\quad$
 \begin{figure}
 	\centering
 	\includegraphics[width=0.98\columnwidth]{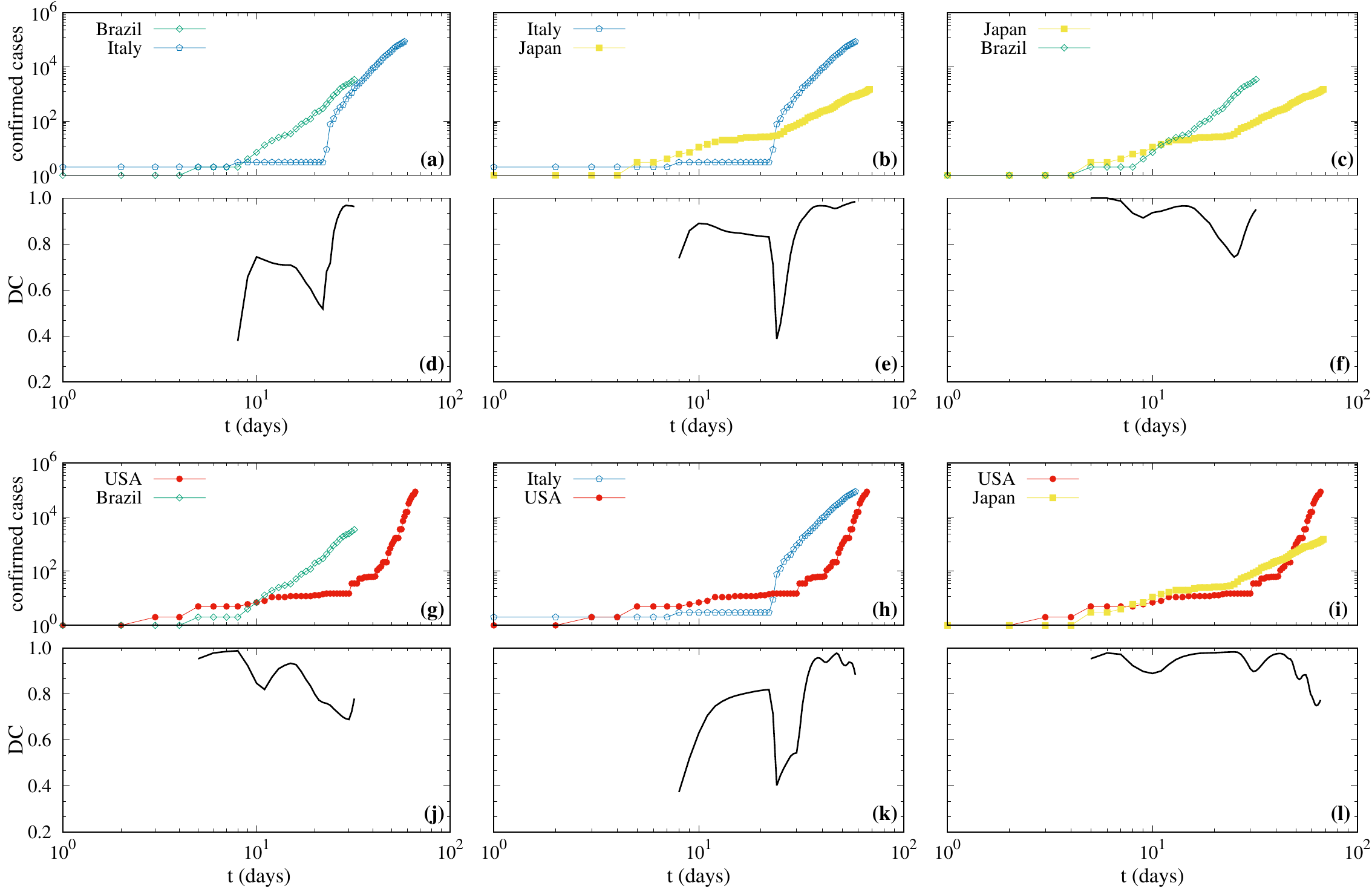}
	\caption{In panels (a), (b), (c), (g), (h), and (i) the log-log plot of the cumulative 
    numbers of confirmed infected cases as a function of time are presented for
    the possible pairs of countries formed between Brazil, Italy, Japan, and
    USA. The semi-log plot of DC calculated between these pairs of countries is 
    presented in panels (d), (e), (f), (j), (k), and (l), respectively.}
 	\label{DC}
 \end{figure}
\end{widetext}

\section{Predictions and strategies}
\label{model}

The model proposed in this work for the numerical prediction and strategies is presented 
in details in Appendix~\ref{apb}. It is a variation of the well known 
Susceptible-Exposed-Infectious-Recovered (SEIR) epidemic model \cite{modeloSeir1, 
modeloSeir2} to propose efficient strategies which allow the government to increase the 
flattening  of the power-law curves. Our SEIR model takes into account the isolation of 
infected individuals \cite{modeloSeiqr1, modeloSeiqr2, modeloSARS, modeloSirx}. In this 
case, the quarantine means the identification and isolation of infected individuals. The 
parameters are divided in two categories: (i) those related to the characteristic of the 
virus spreading, defined {\it a priori} from other studies and (ii) those related to 
adjusting the model to the real data \cbeims{(for more details please see Appendix~\ref{apb})}. 
These parameters can change according to social actions and government strategies.

Numerical results of this section take into account possible interferences or strategies 
from the government of each country, what means that some parameters must be changed after 
the last day of the real data. For each distinct strategy, we use distinct colors which 
are then plotted.
\begin{widetext}
 $\quad$
 \begin{figure}[!t]
	\centering
	\includegraphics[width=0.95\columnwidth]{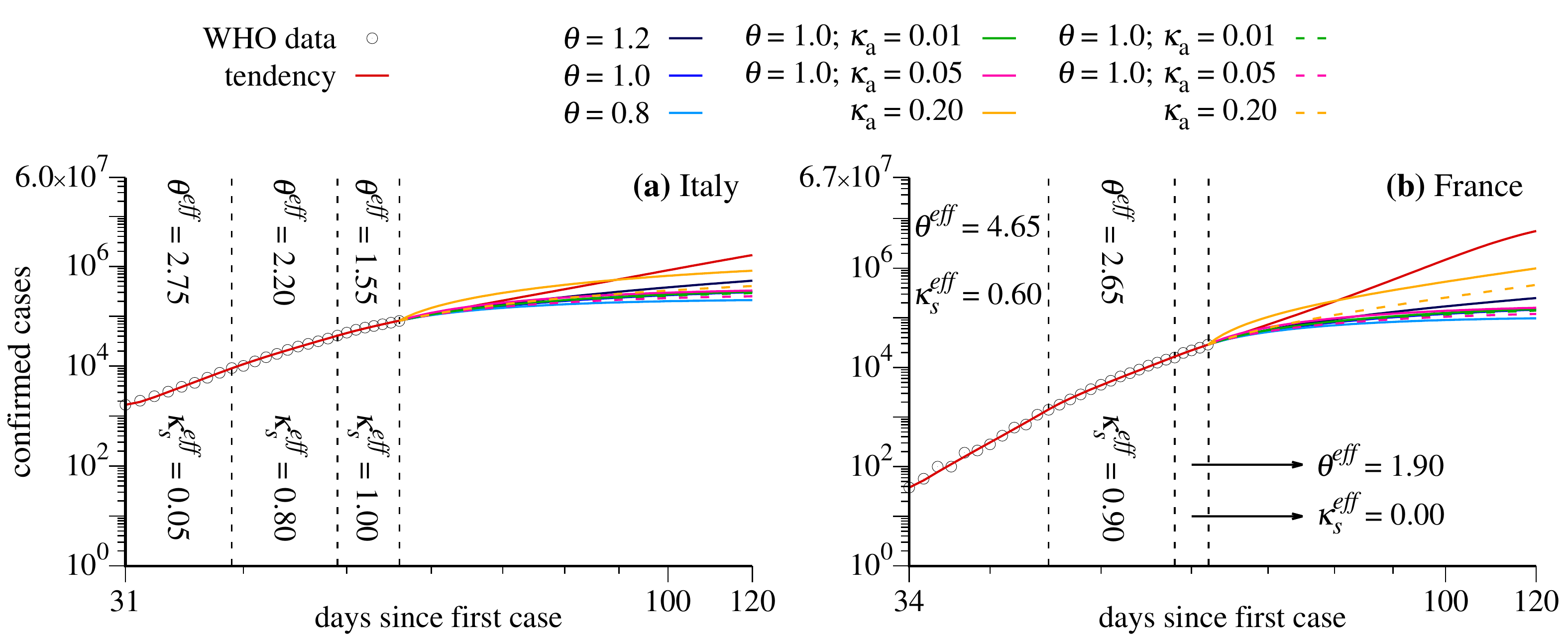}
	\includegraphics[width=0.95\columnwidth]{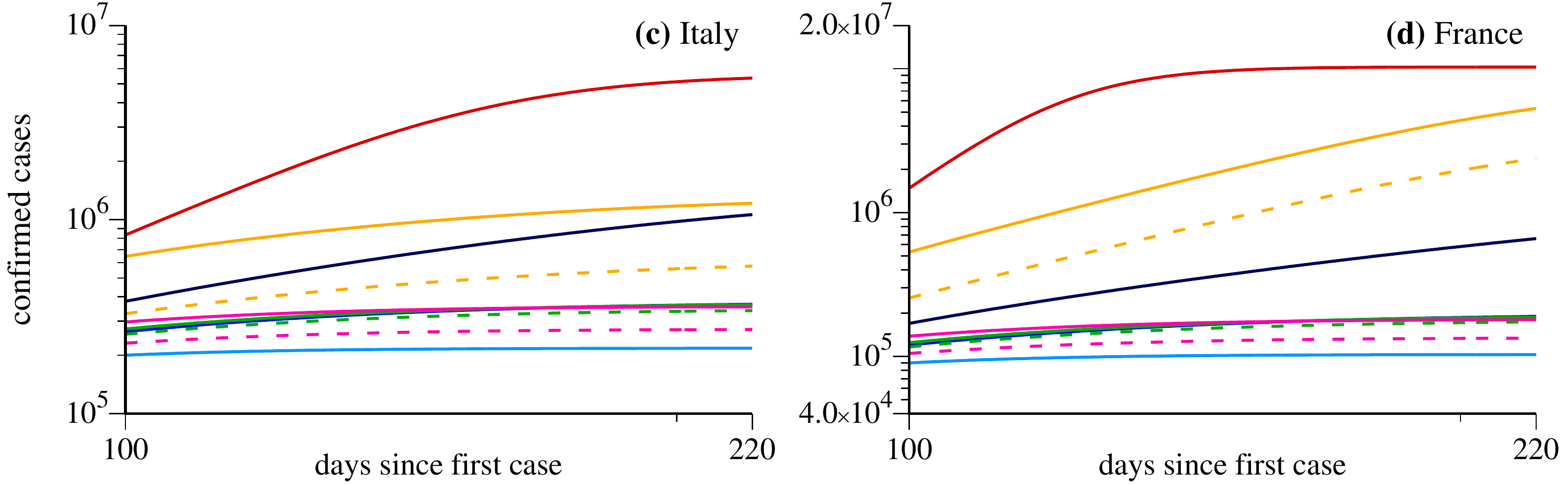}
	\caption{Log-log plot of cumulative number of confirmed cases (black circles) for Italy 
    [(a) and (c)] and France [(b) and (d)] as a function of time and the projected number of 
    cases (colored lines) using distinct government strategies (discussed in the text).}
	\label{france-italy}
 \end{figure}
\end{widetext}
The colors of the subtitles represent the application of distinct 
strategies which leads to distinct {\it scenarios}. For a detailed explanation of variables 
and parameters, please see the Appendix~\ref{apb}. The colors \cbeims{used in Fig.~\ref{france-italy}} 
for the distinct scenarios are the following (for continuous curves):

%\begin{itemize}
%  \item 
\noindent
\xxr{{\bf Red curves}}: the tendency which follows from the behavior of the last points of 
real data (last values for $\theta^{eff}$ and $\kappa_s^{eff}$). This is what happens if 
we do not \cbeims{change the current scenario} on March 
28$^{\mbox{\footnotesize th}}$. \\
%  \item
\xxb{\bf Blue curves}: reduction of social interactions by using smaller values of 
$\theta$. Dark blue for $\theta=1.2$, \cbeims{medium blue for $\theta=1.0$} and light blue 
for $\theta=0.8$. \\
%  \item 
\xxg{\bf Green curves}: reduction of social interaction together with tests to identify 
and isolate \cbeims{asymptomatic and mild symptomatic cases. 
Here we use $\theta=1.0$ and $\kappa_a=0.01$.}\\
%  \item 
{\color{magenta}\bf Magenta curves}: \cbeims{reduction of social interaction together 
with tests to identify and isolate asymptomatic and mild symptomatic cases. Here we use 
$\theta=1.0$ and $\kappa_a=0.05$.}\\
%  \item 
{\color{orange} \bf Orange curves}: \cbeims{identification and isolation of asymptomatic 
and mild symptomatic cases with rate $\kappa_a=0.20$. In this strategy we do not increase 
the social distance and use the last value for $\theta$ obtained in the adjustment.}
  
For the dashed curves the configurations are the same inside each color. However, in these 
curves, the asymptomatic \cbeims{and mild symptomatic identified cases} are not accounted 
for. We notice that without the realization of tests in the population, the asymptomatic 
individuals would not be computed.

We start discussing the cases of Italy and France, shown in Figs.~\ref{france-italy}(a), 
\ref{france-italy}(c) and Figs.~\ref{france-italy}(b), \ref{france-italy}(d), respectively. 
\cbeims{In these cases $\kappa_s=1$, which means that we assume that all symptomatic 
individuals are properly isolated.} Figures \ref{france-italy}(c) and \ref{france-italy}(d) 
\cbeims{show the evolution of scenarios for $120$}. The 
vertical axis is the cumulative number of positive infected individuals in the population. 
In the horizontal axis we have the days since the first computed case in these countries. 
Black circles are the real data starting from the power-law-like behavior discussed in 
Sec.~\ref{data}. During the times for which real data are available, the model chooses the 
values of the parameters $\theta$ and $\kappa_s$ that better adjust the simulation results 
with the data. In the cases shown in Fig.~\ref{france-italy} we needed three values of 
$\theta$ and $\kappa_s$, namely the values $\theta^{eff}$ and $\kappa^{eff}_s$ given in 
the figures. As a consequence, the red curves are in full agreement with the data in this 
time interval. When the available data end, the simulation continues and the red curves 
can be used to predict the asymptotic number of confirmed cases since they represent the 
scenario following the tendency demonstrated by the data. In the case of Italy we obtain 
\cbeims{$5.6\times10^6$ and for France $1.0\times10^7$.} See \cbeims{the tendencies in}
Figs.~\ref{france-italy}(c) and \ref{france-italy}(d). \cbeims{The considerable difference 
between these projections is explained by the last values of $\theta^{eff}$ and 
$\kappa_s^{eff}$ obtained for these countries. Besides $\theta^{eff}$ being larger for France, 
we obtain $\kappa_s^{eff}=0$, which can be interpreted as the nonexistence of quarantine 
measures or the inefficient isolation of symptomatic individuals.} We are aware that such 
asymptotic behavior can be hardly trusted with numerical simulation of models. However, 
our intention in displaying such asymptotic behavior is to show that the proposed model 
converges to reasonable values.

Now we discuss results for some emblematic scenarios for the model when 
\cbeims{specific} strategies are applied to Italy and France on day March 28$^{\mbox{\footnotesize th}}$. 
\cbeims{For both countries we assume $\kappa_s=1$ for all strategies, which means that all 
symptomatic individuals per day are putted into quarantine.} We can see that the strategy 
represented by the orange curves is not sufficient \cbeims{to reduce significantly the total 
number of confirmed infected individuals for France, since the last value $\theta^{eff}=1.90$ 
indicates a large level of social interaction in this country. On the other hand, for Italy a 
considerable reduction is observed, specially for the orange-dashed curve, which indicates only 
the number of symptomatic cases. Strategies related to the blue curves mitigate the growth of 
the number of confirmed cases and, with exception of the dark blue case for Italy, lead to 
smaller asymptotic values when compared to the red curves and orange scenarios. The light 
curves, related to large social distance ($\theta=0.8$), are the most efficient scenarios to 
induce an accentuated reduction of the growth and a fast convergence to the maximal number of 
confirmed cases. Furthermore, green curves tend to approach the medium blue curves, which 
means that, for $\theta=1.0$, there is no significant difference between isolating $1\%$ 
($\kappa_a=0.01$) of the asymptomatic individuals per day or doing nothing. However, increasing 
the daily ratio of detection and isolation of asymptomatic individuals to $\kappa_a=0.05$, a 
noticeable  reduction of the asymptotic value of infected individuals is 
observed (see magenta-dashed curves). Nevertheless, none of these strategies are better than 
increasing the social distance, scenario represented by the light blue curves.}

\begin{widetext}
 $\quad$
  \begin{figure}[!t]
	\centering
	\includegraphics[width=0.92\columnwidth]{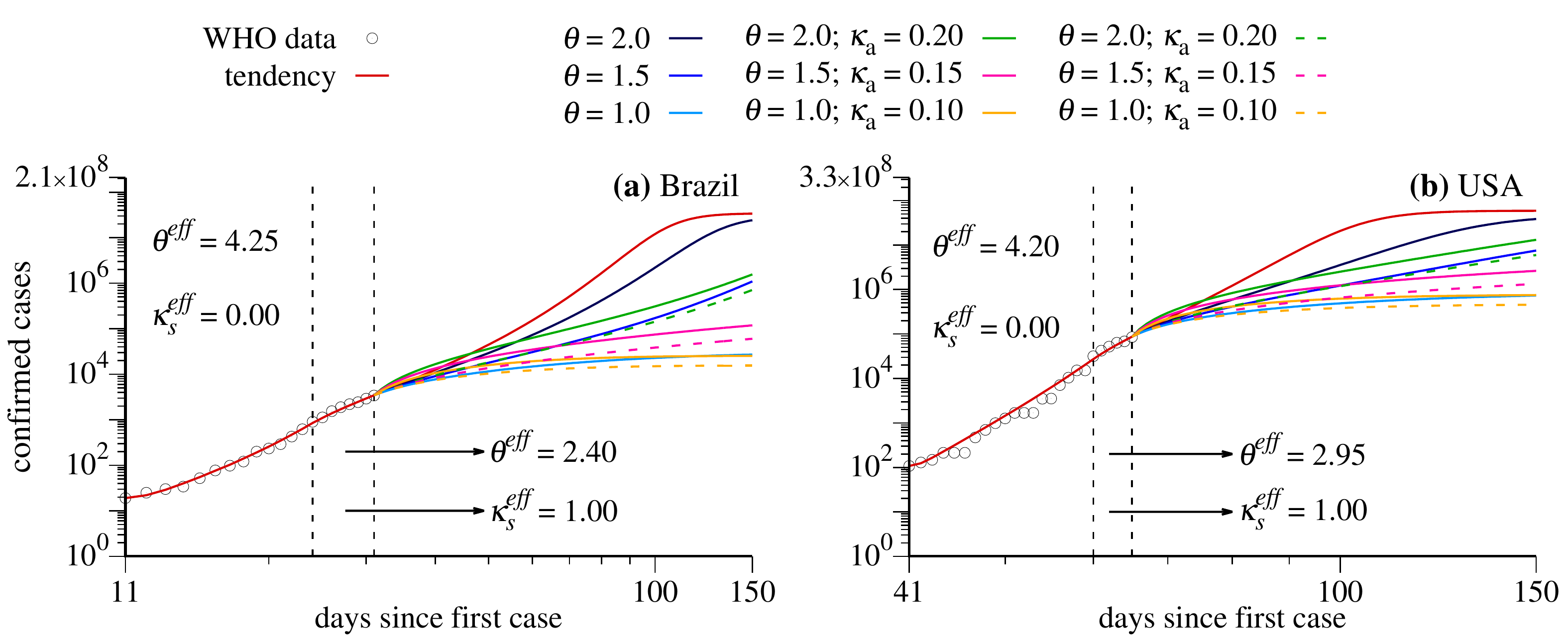}
	\includegraphics[width=0.92\columnwidth]{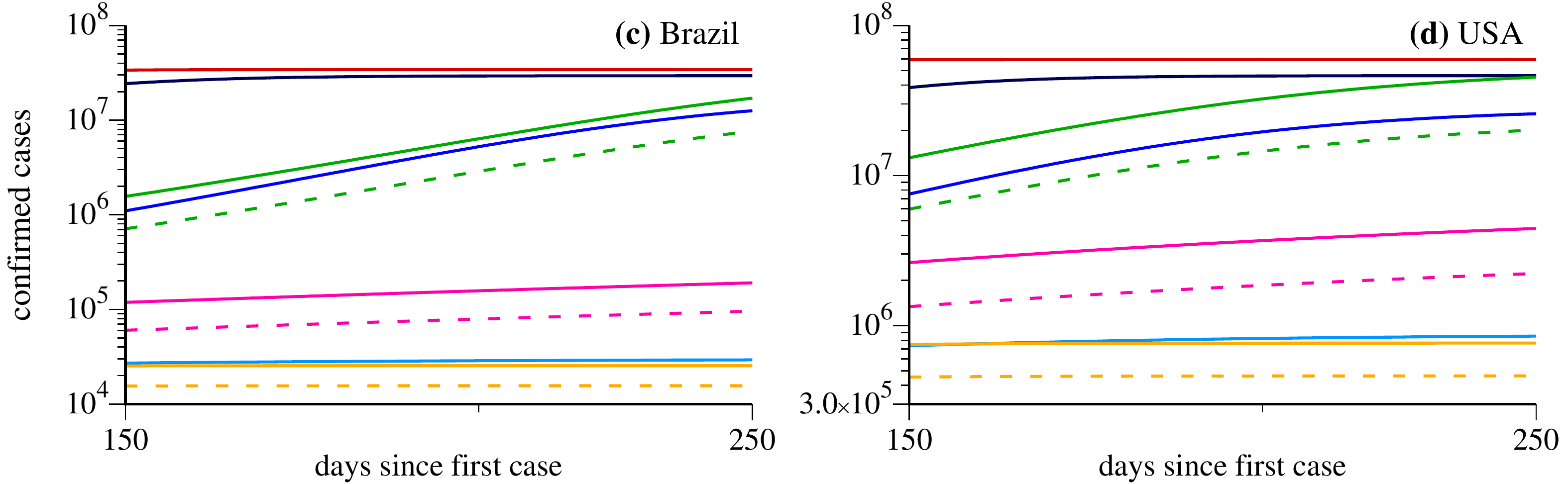}
	\caption{Log-log plot of cumulative number of confirmed cases (black circles) for Brazil 
    [(a) and (c)] and USA [(b) and (d)] as a function of time and the projected number of 
    cases (colored lines) using distinct government strategies (discussed in the text).} 
	\label{brasil-usa}
 \end{figure}
\end{widetext}

At next we discuss cases for Brazil and USA using other strategies. Results are shown 
in Figs.~\ref{brasil-usa}(a), \ref{brasil-usa}(c) and Figs.~\ref{brasil-usa}(b), 
\ref{brasil-usa}(d), respectively. Figures \ref{brasil-usa}(c) and \ref{brasil-usa}(d) 
furnish predictions for the number of infected individuals. Due to the distinct 
scenarios, we had to change a bit the color subtitles (Please see \cbeims{also} 
color labels in Fig.~\ref{brasil-usa}): \\
\noindent
\xxr{\bf Red curves:} have the same meaning as before. \\
\xxb{\bf Blue curves:} are still related to the reduction of social interaction so 
that $\theta$ can take the values $\theta=2.0, 1.5$ and $1.0$, going from dark blue to 
light blue.\\
\xxg{\bf Green curves:} \cbeims{reduction of social interaction together with tests 
to identify and isolate asymptomatic and mild symptomatic cases. Here we use 
$\theta=2.0$ and $\kappa_a=0.20$.}\\
{\color{magenta}\bf Magenta curves}: \cbeims{reduction of social interaction together 
with tests to identify and isolate asymptomatic and mild symptomatic cases. Here we use 
$\theta=1.5$ and $\kappa_a=0.15$.}\\
%  \item 
{\color{orange} \bf Orange curves}: \cbeims{reduction of social interaction together 
with tests to identify and isolate asymptomatic and mild symptomatic cases. Here we use 
$\theta=1.0$ and $\kappa_a=0.10$.}\\
\cbeims{For the dashed curves, the} parameters are the same as those from the continuous 
curves above, but represent the total number of confirmed symptomatic individuals. We 
notice that asymptomatic individuals, or those with very light symptoms, would not be 
identified without realization of tests and are not computed in the number of confirmed 
cases.

As in Fig.~\ref{france-italy}, during the times for which real data are available, the 
model chooses the values of the parameters $\theta$ and $\kappa_s$ that better adjust 
the simulation results with the real data. In the case of Brazil and USA we obtain two 
values of $\theta^{eff}$, as shown in Fig.~\ref{brasil-usa}(a) and (b) with the 
corresponding numerical values. Red curves nicely fit the data as long they are available. 
For the USA case there were some difficult to adjust the parameters since the data show 
some irregularities. In the case for which the strategy does not change ($\theta=\theta^{eff}$ 
and $\kappa_s=\kappa_s^{eff}$), the red curves increase very much for both countries. 
Very high asymptotic values of infected individuals are reached, \cbeims{$3.4\times 10^7$ 
cases for Brazil and $5.9\times 10^7$ cases for USA.}

Now we discuss results when new government strategies are applied to Brazil and
USA on March 28$^{\mbox{\footnotesize th}}$. All blue curves (dark to light) tend to 
\cbeims{mitigate the growth of the asymptotic 
number of cases} for both countries. However, values of $\theta=2.0$ and $1.5$ still 
lead to large asymptotic values. \cbeims{The case with} $\theta=1.0$ is the most relevant 
strategy to flatten the curve efficiently. \cbeims{From the strategies which combine social 
distance with quarantine for the asymptomatic and mild symptomatic cases, stands out the 
inefficiency of the scenario with $\theta=2.0$ and $\kappa_a=0.20$ to flatten the green 
curve. On the other hand, strategies with larger social distance (magenta and orange curves) 
lead to more promising scenarios. The better prediction emerges 
when using $\theta=1.0$ and $\kappa_a=0.10$ (orange curves). For comparison, if we consider 
only the symptomatic cases, we obtain the asymptotic values $3.0\times 10^4$ (light blue 
curve) and $1.6\times 10^4$ (orange-dashed curve) for Brazil and $8.7\times 10^5$ (light blue 
curve) and $4.6\times 10^5$ (orange-dashed curve) for USA. As in the case of France and Italy, 
also for Brazil and USA we observe that the social distance is the key element to reduce the 
growth rate of the maximum of symptomatic cases.} 

%Surprisingly, black and gray curves have strong effects on the flattening, even though 
%$\theta=2.0$. As in Fig.~\ref{france-italy}, dashed curves are very efficient to flatten 
%the curves of both countries.

\section{Conclusions}
\label{con}

The power-law growth of the cumulative number of confirmed infected individuals by the 
COVID-19 until March {27$^{\mbox{\footnotesize th}}$,} 2020, is shown to be the best
description scenario for the countries: Brazil, China, France, Germany, Italy, Japan, 
Spain, Republic of Korea, and USA. Distinct power-law exponents for the countries are 
found and summarized in Table~\ref{T1}. The power-law behavior suggests that the
underlying propagation dynamics of the virus around the countries follows scale free 
networks, fractal kinetics and small world features \cite{watts04,ziff20,ray20}. While 
power-laws with distinct exponents may look similar visually, it is necessary to {\it 
quantify} this similarity. For this we  compute the Distance Correlation \cite{Szek07,
CFMWB,Carlos19} between all countries mentioned above (not shown). However, using 
representative countries of four continents, namely Brazil, Italy, Japan, and USA, 
results for the DC are presented in Fig.~\ref{DC}. They show that the power-law growth
between these countries are highly correlated, even between north and south hemispheres. 
The high correlation between the power-law curves of different continents strongly 
suggest that government strategies can be applied with success around the whole World. 

Furthermore, we propose a variation of the well known SEIR epidemic model 
\cite{modeloSeir1,modeloSeir2} for predictions using (or not) distinct government 
strategies applied on March 28$^{\mbox{\footnotesize th}}$, 2020. We apply numerically 
distinct strategies to flatten (or not) the power-law curves. Even though the social 
isolation, a well know benefit, is very powerful to flatten the curves, we found 
other strategies which lead to comparable results. \cbeims{For Italy and France,
for example, the best scenario was obtained when reducing the social interaction
to $\theta=0.8$ (light blue curves). However, if $5\%$ of the asymptomatic and mild 
symptomatic individuals are identified and isolated every day ($\kappa_a=0.05$),}
\rms{asymptotic values of the same order of magnitude} 
%Valores assintóticos para a Itália: $2.2\times 10^5$ (curva azul claro) e $3.6\times 
%10^5$ (curva magenta tracejada). Valores assintóticos para a França: $1.0\times 
%10^5$(curva azul claro) e $1.8\times 10^5$ (curva magenta tracejada)|} 
\cbeims{for the symptomatic cases were obtained even if $\theta=1.0$ (magenta-dashed curve). 
On the other hand, for Brazil and USA, our simulations confirm that to keep the social 
distance is essential to decrease the asymptotic number of infected individuals, and even 
better results could be obtained by increasing the tests to identify and isolate asymptomatic 
and mild symptomatic individuals (compare light blue and orange curves).} The above 
combination between social interaction and the \cbeims{\it huge} degree of isolation of 
infected  individuals \cbeims{could be implemented} to prevent economic catastrophes 
because people are not working. In other words, let some essential individuals go back to 
work (increasing $\theta$) and, {\it simultaneously}, increase by a  \cbeims{{\it huge}} 
amount  the number of \cbeims{daily} tests and isolation of infected individuals. This 
\cbeims{could furnish} an efficient scenario to flatten the power-law.

\cbeims{Nevertheless, we point out again that our main results confirm that the social 
isolation of individuals is by far the best efficient strategy to flatten the curves.}

\appendix
\numberwithin{equation}{section}

\section{The distance correlation}
\label{apa}

In this section we give a precise definition of the DC following \cite{Szek07}.
Consider  joint random sample $(\boldsymbol{X},\boldsymbol{Y}) = 
\{(X_{k},Y_{k}):k=1,...,N\}$ with $X, Y \in \mathbb{R}^{p}$  and $N \geq 2$, with $i=1,...,N$ 
and $j=1,...,N$. In addition consider the matrix $A_{ij} = a_{ij} - \bar{a}_{i.} - \bar{a}_{.j} 
+ \bar{a}_{..},$ where $a_{ij} = |X_{i} - X_{j}|_{p}$ is the Euclidean norm of the distance 
between the elements of the sample, $\bar{a}_{i.} = \frac{1}{N} \sum_{j=1}^{N} a_{ij}$ and 
$\bar{a}_{.j} = \frac{1}{N} \sum_{i=1}^{N} a_{ij}$ are the arithmetic mean of the rows and 
columns, respectively, and $\bar{a}_{..} = \frac{1}{N^{2}} \sum_{i,j=1}^{N} a_{ij}$ is the 
general mean. A similar matrix $B_{ij} = b_{ij} - \bar{b}_{i.} - \bar{b}_{.j} + \bar{b}_{..},$ 
can be defined using $b_{ij} = |Y_{i} - Y_{j}|_{p}$. The terms $b_{ij}$, $\bar{b}_{i.}$, 
$\bar{b}_{.j}$ and $\bar{b}_{..}$ are similar to those from matrix $A_{ij}$. From these 
matrices we compute the {\it empirical distance correlation} from 
\begin{equation}\label{dc}
DC_{N}(\boldsymbol{X},\boldsymbol{Y})=\frac{\sigma_{N}(\boldsymbol{X},\boldsymbol{Y})}{
\sqrt{\sigma_{N}(\boldsymbol{X})\sigma_{N}(\boldsymbol{Y})}},
\end{equation}
where 
\begin{equation}\label{} 
\sigma_{N}(\boldsymbol{X},\boldsymbol{Y}) = \frac{1}{N}\sqrt{ \sum_{i,j=1}^{N} A_{ij}B_{ij}},
\end{equation}
and

\begin{equation}\label{} 
\sigma_{N}(\boldsymbol{X}) = \frac{1}{N}\sqrt{\sum_{i,j=1}^{N} A^{2}_{ij}},
\quad
\sigma_{N}(\boldsymbol{Y}) = \frac{1}{N}\sqrt{\sum_{i,j=1}^{N} B^{2}_{ij}}.
\end{equation}

\section{The Model}
\label{apb}

The proposed model contains six ODEs and is an extension of a model known in the 
literature \cite{modeloSeiqr1,modeloSeiqr2,modeloSARS}. Many other related models have been 
proposed with distinct characteristics 
\cite{modeloSirx,modeloShir,modeloSiqr,modeloSeiqr1,modeloSeiqr2,modeloSARS,modeloSeir2}. In 
our case we consider symptomatic $I_s$ and asymptomatic $I_a$ infected
individuals \cbeims{($I_a$ also includes individuals with mild symptoms).} 
%\edu{Seria com sintomas não leves em $I_s$ e assintomáticos e casos leves em $I_a$}. 
Quarantine $Q$ is also contemplated, respectively. The transition rate from asymptomatic to 
symptomatic cases is neglected as an first approach. The Ordinary Differential Equations (ODEs) 
are given by
\begin{align}
	\dot{S}   &= -\frac{\theta}{T_{inf}}\frac{(I_s + \alpha I_a)}{N}S,\\
	\dot{E}   &= \frac{\theta}{T_{inf}}\frac{(I_s + \alpha I_a)}{N}S - \frac{E}{T_{lat}},\\
	\dot{I}_s &= (1 - \beta)\frac{E}{T_{lat}} - \left(\kappa_s + \frac{1}{T_{inf}}\right)I_s,\\
	\dot{I}_a &= \beta\frac{E}{T_{lat}} - \left(\kappa_a +  \frac{1}{T_{inf}}\right)I_a,\\
	\dot{Q}   &= \kappa_s I_s + \kappa_a I_a - \frac{Q}{T_{serial}},\label{quarentena}\\
	\dot{R}   &= \frac{I_s + I_a}{T_{inf}} + \frac{Q}{T_{serial}}.
%	\dot{R}_i &= \frac{1}{T_{inf}}\left(I_s + Q \right),\\
%	\dot{R}_a &= \frac{I_a}{T_{inf}}.
\end{align}
The dot represents the time derivative and the variables are:
\begin{itemize}
	\item $N = S + E + I_s + I_a + Q + R$: total population. 
	\item $S$:  individuals susceptible to infection.
	\item $E$:  exposed individuals, remain latent until infected.
	\item $I_s$: symptomatic individuals. \cbeims{Represent individuals
	 with strong symptoms. We assume that these individuals look for health 
	 care and are included in the confirmed cases.}  
	\item $I_a$: asymptomatic individuals \cbeims{and mild symptomatic cases.} 
	\item $Q$: infected individuals isolated (in quarantine),
	\item $R$: individuals which were infected and not identified but became immune.
\end{itemize}
To adjust the parameters following distinct countries, as well as government 
measures, we used the cumulative number of confirmed infected individuals
$C_{\rm{cum}}$.The number of confirmed infected individuals as a function of time 
is defined by $C(t)=I_s(t)+Q(t)$. Parameters which do not depend on strategies are:
\begin{itemize}
	\item \cbeims{$T_{serial}=7.5$ days: mean serial interval~\cite{tempoSerialIncubacao}. 
    Is the mean time between successive cases of the transmission of the disease}.
%    \edu{Tempo médio entre casos sucessivos da transmissão da doença.} 
%    Seria necessário uma referência para esse significado? Posso procurar algo.}
	\item $T_{lat}=5.2$ days: \cbeims{mean incubation period~\cite{tempoSerialIncubacao}. 
    \cbeims{Assumed to be equal to the latent time}~\cite{modeloSeir2},} which is the time 
    interval the exposed individual \cbeims{remains infectious}, also denominated incubation 
    time.
	\item $T_{inf}=T_{serial}-T_{lat}$: \cbeims{infectious period}~\cite{modeloSeir2}.
	\item $\alpha=1.0$: ratio between infectiousness of asymptomatic and symptomatic 
    individuals. \cbeims{We assume 
    that the numbers of asymptomatic and symptomatic individuals are equal.}
	\item $\beta=0.8$: population ratio which remains asymptomatic \cbeims{or mild symptomatic, 
    which is the most common observed cases (see the situation report $46$ in Ref.~\cite{OMS}).}
\end{itemize}
Parameters which are related to the use of distinct strategies are:
\begin{itemize}
   \item $\theta=\gamma R_0$: replication factor, with $\gamma$ being
   a number that represents the proportion of interaction between individuals and 
   $R_0$ the basic reproduction number. In our model, $\theta$ is an adjustable 
   parameter according to WHO data. 
    \item $\kappa _s$: \cbeims{isolation rate of symptomatic individuals.}
	\item $\kappa _a$: \cbeims{isolation and identification rate of asymptomatic individuals.} 
\end{itemize}
In this model, no rigid quarantine is taken into account and no immunization
term is defined, since until today no vaccine has been developed. In 
Eq.~(\ref{quarentena}), the factor $T_{serial}$ dividing $Q$ represents a rate of 
exit from the quarantine (for the group $R$).

It is \cbeims{of most relevance}  to mention that the \cbeims{only adjustable 
parameters in our simulations are} $\theta$ and $\kappa_s$. \cbeims{This is
important since for systems composed of  differential equations with 
$r$ parameters, $2r + 1$ experiments are need to obtain all the information that 
is potentially available about the parameters \cite{son02}. Since in our case $r=2$, 
we need $5$ real data to adjusted parameters correctly. All real data used in 
Figs.~\ref{france-italy} and \ref{brasil-usa} to find $\theta^{eff}$ and $\kappa^{eff}_s$
are larger than $5$. Furthermore, the initial condition for the variable $E(t_0)$ is 
adjusted only in the first part of the data, where the first $\theta^{eff}$ and 
$\kappa^{eff}_s$ are determined along the data. It could be thought that such initial 
condition must be include in the adjustable parameters. However, even in such case, 
$r=3$ and $2r+1=7$.  The lowest number of available data in the first part of data is 
$9$, as can be seen in Fig.~\ref{france-italy}(b), meaning that even in the worst case 
our adjustments are trustful.} The goal is to minimize the mean square error between 
the predicted curve and real data. \cbeims{In the case analyzed here, $\kappa_a=0$ along 
the real data. It is only changed, not adjusted,} when there are information available 
about the test realization in the population. We do not start the parameter's adjustment 
from the first day of reported infections, but later on. The model produces better 
results in such cases.

\cbeims{Regarding the adjustment of the parameters $(\theta,\kappa_s)=(\theta^{eff},\kappa_s^{eff})$, 
we minimize the mean square error separately inside the three set of data in 
Figs.~\ref{france-italy}(a) and \ref{france-italy}(b), and inside the two set of data in 
Figs.~\ref{brasil-usa}(a) and \ref{brasil-usa}(b). To do so we vary the parameters inside the 
intervals $\theta \in [0,6]$ and $\kappa_s \in [0,1]$ with a step $0.05$.
The initial condition for $E(t_0)$ is determined inside the first part of the data 
considering $(\theta,\kappa_s,E(t_0))$ in the interval
$E(t_0) \in [0,5000]$ using a step equal $5$.}

The initial condition set used in the numerical integration process of
the ODEs is the following: $Q(t_0)=R(t_0)=0$; $I_a(t_0)=\beta I_s(t_0)/(1-\beta)$,
where $I_s(t_0)$ represent the number of confirmed infected cases,
obtained from the real time-series  for the day $t_0$. $S(t)$ is
determined accordingly to the total number of people for each country
and the previous initial conditions. 

\vspace*{1cm}
\acknowledgments{The authors thank CNPq (Brazil) for financial support (grant numbers 
432029/2016-8, 304918/2017-2, 310792/2018-5 and 424803/2018-6) and, they also 
acknowledge computational support from Prof.~C. M. de  Carvalho at LFTC-DFis-UFPR 
(Brazil). C. M. also thanks FAPESC (Brazilian agency) for financial support and 
Luiz A. F.~Coelho and Joel  M.~Crichigno Filho for fruitful discussions.}

\section*{Data availability}
The data that support the findings of this study are openly available
in WHO (World Health Organization), situation reports 1--68 \cite{OMS}.

%\bibliography{references}


\end{document}